\documentclass[letter,singlecolumn]{scrartcl}

\usepackage[T1]{fontenc}
\usepackage{amsmath} 
\usepackage{amssymb}  
\usepackage{xcolor,mathrsfs}
\usepackage{%
	algorithm,%
    algpseudocode,
	amsthm,%
	bbm,%
    booktabs,%
	enumerate,%
	hyperref,%
	mathtools,%
	paralist,%
	pgf,%
	pgfplots,%
	subfigure,%
	tikz,%
	tikzscale,%
}

\algnewcommand{\TRUE}{\textbf{true}}
\algnewcommand{\FALSE}{\textbf{false}}
\usepackage{pgfplots}
\usepgfplotslibrary{fillbetween}
\usetikzlibrary{patterns}
\newcommand{\uv}{\textbf{u}}
\newcommand{\vu}{\textbf{v}}
\newcommand{\w}{\textbf{w}}
\newcommand{\x}{\textbf{x}}
\newcommand{\y}{\textbf{y}}
\newcommand{\z}{\textbf{z}}

\newcommand{\q}{\textbf{q}}

\newcommand{\e}{\lambda}
\newcommand{\J}{\mathcal{J}}
\newcommand{\G}{\mathcal{G}}

\newtheorem{thm}{Theorem}

\newtheorem{cor}[thm]{Corollary}

\newtheorem{lem}[thm]{Lemma}
\newtheorem{asu}{Assumption}
\newtheorem{defn}{Definition}

\begin{document}

\title{Learning Global Equilibria in Non-Concave Prospect Theoretic Games\footnote{This work was supported by D\'efi Inria-EDF}}  
\date{}
\author{Ashok Krishnan K.S.\footnote{Inria and DI ENS, Paris, France. \emph{Email:~ashok-krishnan.komalan-sindhu@inria.fr}}, Lilou Malfoy\footnote{ENSIIE, France. \emph{lilou.malfoy12@gmail.com}}, H\'el\`ene {Le Cadre}\footnote{Inria, Lille, France. \emph{Email:~helene.le-cadre@inria.fr}}, Ana Bu\v si\'c\footnote{Inria and DI ENS, Paris, France. \emph{Email:~ana.busic@inria.fr}}}

\maketitle

\begin{abstract}
Human agents in non cooperative games often exhibit non rational behaviour. Prospect theory provides a behavioural framework to capture how agents perceive gains and losses under risk, but its incorporation into games typically leads to non-concave and non-convex utility functions, making equilibrium computation challenging. In such cases, it is difficult to obtain algorithms that have global convergence behaviour. This paper proposes a learning algorithm  which uses minorization-maximization (MM). Under a proximal Polyak–\L{}ojasiewicz (P\L{}) condition on the potential function, we establish global convergence of the algorithm to a potential-maximizing Nash equilibrium. We further show how the structure of the MM minorizer can be used to enable a distributed implementation of the algorithm. With additional noise injection, the individual updates of the agents can be obscured, while still ensuring global convergence.
\end{abstract}

\section{Introduction}
Non cooperative games have been used successfully to model various problems that arise in areas such as engineering and economics. The classical framework of expected utility maximization has been used predominantly in these works. Usually, these generally assume fully rational agents. However, human agents exhibit a number of non rational behaviours, which reflect in their game play and decision making. Consequently, there have been a number of different approaches that sought to incorporate bounded rational behaviour of human agents in games. Prospect theory (PT) emerged as a framework for capturing systematic departures from expected utility maximization in human decision-making under risk.

A number of papers have used PT in game theory. However, the utility model under PT tends to be in general non-concave and non-smooth. This makes it difficult to directly apply a number of existing game theoretic tools and learning algorithms. In this paper, we consider a non cooperative game model, with agents taking actions from continuous, compact sets. The game is a potential game, and the utility is assumed to be continuously differentiable. However, no assumption is made on the concavity (or convexity) of the utility function (and consequently, the potential function). A learning approach based on minorization-maximization (MM) is proposed. It is shown that the minorization functions can be chosen such that the MM learning approach converges to a potential maximizing Nash equilibrium. The result hinges on the potential function obeying a proximal Polyak-\L{}ojasiewicz (P\L{}) inequality, which ensures that stationary points over the strategy set are maximizers of the potential.

In many applications, agents may be reluctant to share information about their personal preferences with other agents. Prospect-theoretic value functions encode such preferences by capturing how agents perceive and evaluate outcomes under risk. In this paper, we provide an MM algorithm, that can be implemented in a distributed fashion, with noise injected to limit the disclosure of their private preferences. Thus, in non-concave games, it is possible to develop learning algorithms that are distributed and still have global convergence guarantees.

\subsection{Literature Review}
Prospect theory~\cite{kai1979prospect} and its cumulative version~\cite{tversky1992advances} have been incorporated in non-cooperative games in several ways. Equilibrium notions for agents with PT and CPT preferences, and their existence and properties, have been studied in~\cite{keskin2016equilibrium,metzger2019noncooperative,phade2019geometry}, which show that probability weighting can substantially change the structure of the equilibrium set. Learning in repeated games with CPT agents was considered in~\cite{phade2023learning}, where calibrated learning is shown to converge to a set of CPT correlated equilibria. PT has also been used in engineering applications, for instance to model prosumers in smart grid energy management~\cite{etesami2018stochastic}. These works mostly consider finite action sets or specific game structures; in contrast, we consider continuous action sets and study learning dynamics with global convergence guarantees to Nash equilibria. 

In continuous games, convergence of learning dynamics is typically established under concavity or monotonicity of the utilities~\cite{ks2025irrationality,ashok2025achieving}. Without such assumptions, gradient-based learning may fail to converge, or may avoid some Nash equilibria~\cite{mazumdar2020gradient}. In \cite{fochesato2025noncooperative}, the authors propose the use of local Nash equilibria to study non-concave, non-smooth games. Potential games~\cite{monderer1996potential} have the useful property that learning in the game can be related to the optimization of the potential function~\cite{marden2009cooperative}; however, when the potential is non-concave, only convergence to stationary points can be expected in general. The Polyak-{\L}ojasiewicz (P\L{}) inequality~\cite{polyak1963gradient} provides a way of obtaining global convergence without concavity, with linear rates for gradient and proximal-gradient methods under its proximal version~\cite{karimi2016linearArxiv}. In games, a two-sided P\L{} condition has been used to obtain global convergence of gradient methods in non convex-non concave minimax problems~\cite{yang2020global}. Here, we use the proximal P\L{} inequality on the potential function of a game with PT utilities.

MM algorithms~\cite{hunter2004tutorial,sun2017majorization} successively optimize surrogate functions that bound the objective and are tight at the current iterate. For non convex problems, their convergence to stationary points has been established under general conditions~\cite{razaviyayn2013unified,bolte2016majorization}. In multi-agent problems, surrogates obtained by partially linearizing the objective lead to decompositions that can be solved in parallel by the agents~\cite{scutari2014decomposition}. Our minorizer follows a similar idea: the coupling term is kept exact, while the individual PT terms, which encode private preferences, are linearized.

Distributed algorithms for computing Nash equilibria based on the alternating direction method of multipliers (ADMM) have been proposed in~\cite{salehisadaghiani2019distributed}. When the agents' objectives are private, a common approach is to perturb the exchanged messages with noise, which leads to a trade-off between privacy and accuracy~\cite{huang2015differentially}; for such message perturbation schemes, exact convergence and privacy guarantees cannot be achieved simultaneously~\cite{nozari2018differentially}. In this paper, the noisy ADMM is analyzed as an inexact maximization step within the MM scheme, which yields convergence guarantees when the noise variance decreases.

The main contributions of this paper are as follows.
\begin{enumerate}
    \item We propose an MM learning scheme for non-cooperative potential games with prospect theoretic, non concave potentials. We show that, when the potential
    satisfies a proximal P\L{} inequality, the scheme converges at a linear rate to a potential maximizing Nash equilibrium.
    \item We construct a separable  minorizer that allows the maximization step to be carried out in a distributed fashion using ADMM, in which each agent's preferences enter only its own
    local update.
    \item We analyze MM with an inexact maximization step, and apply this
    analysis to a noisy ADMM in which agents perturb their updates to obscure their preferences. We show that the iterates converge when the noise variance decreases, up to an error that decays exponentially in the number of ADMM iterations.
\end{enumerate}

\section{Overview of Prospect Theory}
We now present a high level overview of Prospect theory~ \cite{kai1979prospect},~\cite{stott2006cumulative}. Under the expected utility paradigm, a user evaluates their \emph{prospects} based on expected utility. A prospect $P$ involves outcomes $(x_1,x_2,..,x_M)$ with a distribution $(q_1,...,q_M)$,  where the outcomes are ordered in increasing order of attractiveness. The expected utility is 
\begin{align}
    E(P)\triangleq\sum_{m=1}^Mq_mx_m.
\end{align}
Comparing $P$ to another prospect $\tilde P$ having outcomes $(\tilde x_1,...,\tilde x_M)$ with distribution $(\tilde q_1,...,\tilde q_M)$, involves a comparison between $E(P)$ and $E(\tilde P)$, and choosing the higher one.

\begin{figure}
    \centering
    \begin{tikzpicture}[scale=1]
 
  \draw[gray!20, very thin] (-4,-2) grid (4,2);
 
  \draw[->, thick] (-4.3,0) -- (4.3,0);
  \draw[->, thick] (0,-2.3) -- (0,2.3) node[above] {value};

 \node at (3,0.3) {outcome};
  \foreach \x in {-4,-3,-2,-1,1,2,3,4}
    \draw (\x,0.08) -- (\x,-0.08) node[below] {\footnotesize $\x$};
  \foreach \y in {-2,-1,1,2}
    \draw (0.08,\y) -- (-0.08,\y) node[left] {\footnotesize $\y$};
  \node[below left] at (0,0) {\footnotesize $0$};

  \draw[very thick, blue, domain=-4:4, samples=200, smooth]
    plot (\x, {3.6/(1+exp(-1.5*\x)) - 1.8});
 
\end{tikzpicture}
    \caption{Example of a prospect theoretic value function $v$ that maps outcomes to values.}
    \label{fig:distortion-curve-val}
\end{figure}
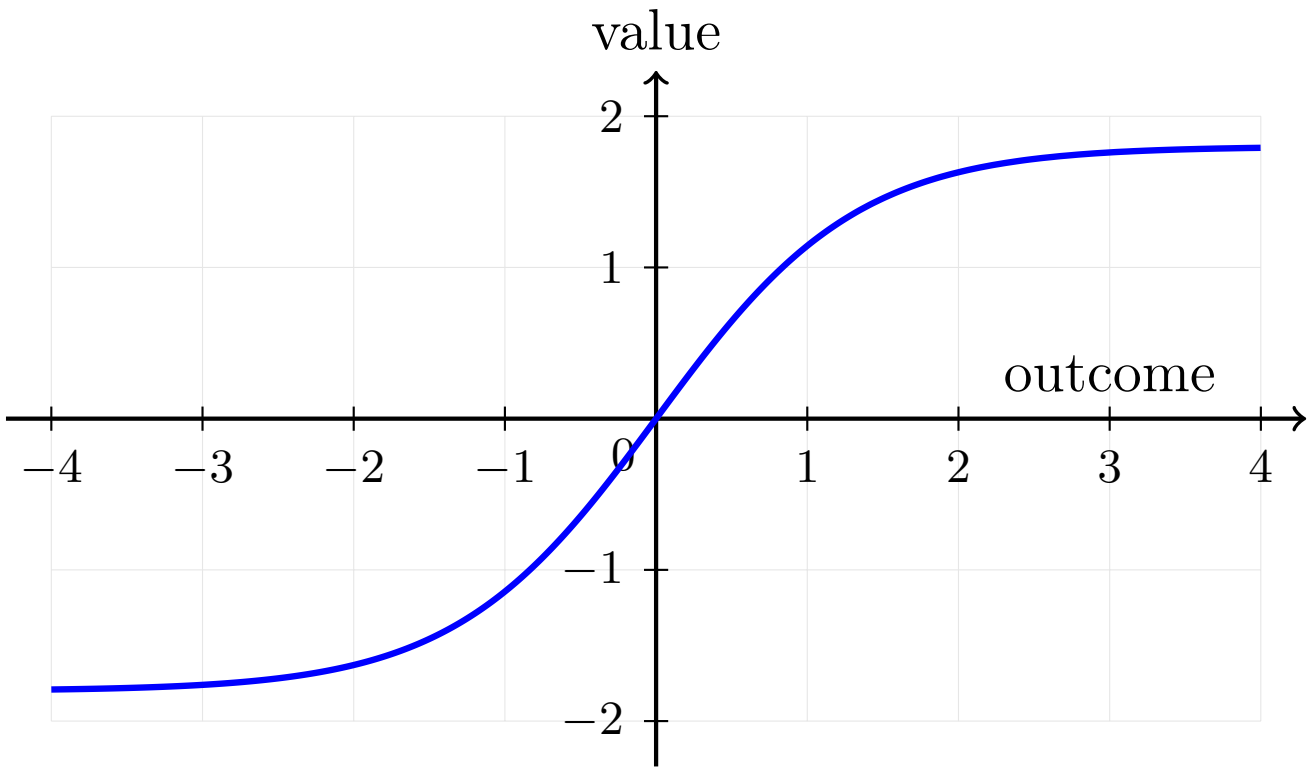

 Under PT, the perception of outcomes changes with respect to a reference point. Outcomes above this reference are seen as gains, and those below as  losses. When faced with gains, agents will behave in a \emph{risk averse} manner, and when faced with losses, agents tend to be \emph{risk seeking} or \emph{risk neutral}. This perception is captured by an outcome to value map $v$, which, in order to model the above (and other) aspects of bounded rational behaviour, is assumed in general to be convex below a threshold, concave above it.  One such function is depicted in Fig. \ref{fig:distortion-curve-val}. Here the agents are risk averse for gains and risk seeking for losses, with the  zero being the reference point that demarcates gains from losses.

In addition to the notion of value relative to risk, the weight accorded to outcomes is modified as well. The modified value of the prospect is
$$
    V(P)\triangleq\sum_{m=1}^M \sigma_mv(x_m),
$$
where $v$ is a function similar to the one in  Fig.~\ref{fig:distortion-curve-val}, and $\sigma_m$ are the weights for the distorted outcomes, given by~\cite{stott2006cumulative},
\begin{align}\label{eqn:q-tilde-gen}
    \sigma_1 =\pi(q_1),
    \sigma_j = \pi(\sum_{m=1}^j q_m)-\pi(\sum_{m=1}^{j-1} q_m),~j>1,
\end{align}
for an appropriate weighting function $\pi$, which is monotone increasing and bounded between 0 and 1. This models  the overweighting of small probabilities, and the underweighting of large probabilities, for large outcomes. Faced with a choice between two prospects $P$ and $\tilde P$, the agent will make a decision comparing their PT modified values $V(P)$ and $V(\tilde P)$, and choosing the better option. It is clear that these decisions are likely to be very different from those predicted by expected utility theory.

\section{Game Model}\label{sec:game-model}
\noindent We consider a game with $N+1$ agents, of which $N$ play the game, while agent $N+1$ acts as a coordinator between the agents. Let $\mathcal N\triangleq\{1,...,N\}$ denote the set of agents that participate in the game. Agent $i\in\mathcal N$ chooses its strategy $x_i$ from a convex and compact set $\mathcal X_i\subseteq\mathbb R$. Let $\x\triangleq(x_i)_{i\in\mathcal N}$ denote the joint strategy vector, which is an element of  $\mathcal X\triangleq\prod_{j=1}^N \mathcal X_j$. The randomness in outcomes is modelled by a real valued random variable $\xi$ taking values over the set  $\Xi\triangleq\{\xi_1,...,\xi_M\}$\footnote{Most of the results in this work can be extended to the case where $\xi$ has continuous distribution, albeit with more conditions on other variables to ensure integrability. We restrict ourselves to a finite support for simplicity.} where $0<\xi_1<\cdots <\xi_M$, with probability distribution $\q=(q_1,...,q_M)$. Let $V_i:\mathbb R\to\mathbb R,i=1,...,N$ and $\pi:[0,1]\to[0,1]$ be monotone increasing. The utility for user $i$  under the strategy $\x=(x_i,\x_{-i})$ where  $\x_{-i}\triangleq(x_1,..,x_{i-1},x_{i+1},...,x_N)$, is given by
	\begin{align}\label{eq:utility-fn-def}
		J_i(x_i,\x_{-i}) \triangleq a_i\J(\x)+\mathbb{E}_{\sigma} [V_i\circ\mathcal R_i(x_i,\xi)],
	\end{align}
	where $0<a_N\le a_{N-1}\le ...\le a_1$, 
	and $\sigma$ is the distribution generated from $\q$ by \eqref{eqn:q-tilde-gen} using $\pi$. This form of the utility function includes the following terms:
	\begin{enumerate}
		\item A  collective   usage benefit $\J(\x):\mathcal X\to\mathbb R$, which is a  collective utility common to all agents.
		\item An individual random reward $\mathcal R_i(x_i,\xi):\mathcal X_i\times\Xi\to\mathbb R$. This reward is random due to its dependence on $\xi$, and is viewed subjectively through the PT value function $V_i$.
	\end{enumerate}
We assume that the functions $\J$, $V_i$ and $\mathcal R_i$ are continuous and  smooth, for each $i$.
\begin{defn}
        We define the game $\G$ as the tuple $\left(\mathcal N,\{\mathcal X_j\}_{j=1}^{N},\{J_j\}_{j=1}^{N},\{V_j\}_{j=1}^N,\sigma\right)$.
    \end{defn} 
\begin{defn}
    A Nash equilibrium for $\G$ is any $\x^*\triangleq(x^*_i,\x_{-i}^*)=(x_1^*,...,x_N^*)\in\mathcal X$ such that for all $i=1,...,N$, 
	\begin{align}\label{eq:def-nash-eq}
		J_i(x^*_i,\x^*_{-i})\ge J_i(x,\x^*_{-i}),~\forall x\in\mathcal X_i.
	\end{align}
\end{defn}
\subsection*{Potential Structure}
The game $\G$ is a weighted potential game~\cite{monderer1996potential}, with weights $a_i$ and potential function
\begin{align}
    \Phi(\x)\triangleq \J(\x)+\sum_{j=1}^N\frac{1}{a_j}\mathbb E_{\sigma} [V_j\circ\mathcal R_j(x_j,\xi)],
\end{align}
that is, $J_i(x_i,\x_{-i})-J_i(\tilde x_i,\x_{-i})=a_i(\Phi(x_i,\x_{-i})-\Phi(\tilde x_i,\x_{-i}))$ for all $i$, $\x\in\mathcal X$ and $\tilde x_i\in\mathcal X_i$. Since $\Phi$ is continuous and $\mathcal X$ is compact, the set of maxima
\begin{align}
    \mathbf M\triangleq \arg_{\x\in\mathcal X}\max \Phi(\x)
\end{align}
is non empty, and every element of $\mathbf M$ is a Nash equilibrium~\cite{monderer1996potential}; we call such points \emph{potential optimal equilibria}.
Define the maximum value of the potential function and the optimality gap,
\begin{align}
    \Phi^*=\Phi(\x)~\text{for some~}\x\in\mathbf M,\quad g(\x)\triangleq\Phi^*-\Phi(\x).
\end{align}

We make the following assumption.
\begin{asu}[Proximal P\L{} Inequality]\label{asu:PL-condition}
    There exist $\mu>0$ and $\bar L>0$ such that the potential function satisfies, for all $\x\in\mathcal X$,
    \begin{align}\label{eqn:prox-PL}
        \frac{1}{2}\mathcal D_{\bar L}(\x)\ge \mu g(\x),
    \end{align}
    where, for $L>0$,
    \begin{align}\label{eqn:prox-PL-D}
        \mathcal D_L(\x)\triangleq 2L\max_{\y\in\mathcal X}\left(\langle\nabla\Phi(\x),\y-\x\rangle-\frac{L}{2}||\y-\x||^2\right).
    \end{align}
\end{asu}
Assumption \ref{asu:PL-condition} is a generalization of the P\L{} inequality to constrained problems~\cite{karimi2016linearArxiv}. The maximum in \eqref{eqn:prox-PL-D} is attained at  $\y=\Pi_{\mathcal X}\left(\x+\frac{1}{L}\nabla\Phi(\x)\right)$, where $\Pi_{\mathcal X}$ denotes projection to the set $\mathcal X$. The function $\mathcal D_L(\x)$ satisfies $\mathcal D_L(\x)\ge 0$, with equality if and only if $\x$ is a stationary point of $\Phi$ over $\mathcal X$, i.e., $\langle\nabla\Phi(\x),\y-\x\rangle\le 0$ for all $\y\in\mathcal X$.  Whenever the gradient step $\x+\frac{1}{L}\nabla\Phi(\x)$ lies in $\mathcal X$ (in particular, if the constraints were absent), $\mathcal D_L(\x)=||\nabla\Phi(\x)||^2$, and \eqref{eqn:prox-PL} reduces to the classical P\L{} inequality
$
    \frac{1}{2}||\nabla\Phi(\x)||^2\ge \mu g(\x).
$
The classical P\L{} inequality ensures that all points with zero gradient are optimal points for unconstrained optimization. 
The proximal P\L{} assumption ensures that every stationary point of $\Phi$ over $\mathcal X$, including stationary points on the boundary of $\mathcal X$, is also a maximizer. Thus, learning algorithms that move towards a stationary point are assured of reaching an optimum.

The following property of the function $\mathcal D_L$ will be used in subsequent results.
\begin{lem}\label{lem:D-increasing-property}
    For $L_1\ge L_2>0$ and all $\x\in\mathcal X$,
    \begin{align}
        \mathcal D_{L_1}(\x)\ge\mathcal D_{L_2}(\x).
    \end{align}
\end{lem}
\begin{proof}
    \begin{align}
        \mathcal D_L(\x)
         &= \max_{\y\in\mathcal X}\left(2L\langle\nabla\Phi(\x),\y-\x\rangle-L^2||\y-\x||^2\right),\\
         &=\max_{\z\in L(\mathcal X-\x)}\left(2\langle\nabla\Phi(\x),\z\rangle-||\z||^2\right).
    \end{align}
    The result follows by observing that any element in the set $L_2(\mathcal X-\x)$ is contained in the set $L_1(\mathcal X-\x)$.
\end{proof}

\section{Convergence under the Proximal P\L{} Condition}\label{sec:PL-conv}
\begin{defn}[MM Learning] Consider the sequence of joint strategies $(\x(k),k\ge 0)$, where 
    \begin{align}
    \x(k+1)=\arg_{\x\in\mathcal X}\max\phi(\x|\x(k)),
    \end{align}
    where $\phi(\cdot|\x(k))$ is the minorizing function at iteration $k$, which is any function that satisfies,
    \begin{enumerate}
        \item $\phi(\x|\x(k))\le \Phi (\x),~\forall\x\in\mathcal X$,
        \item  $\phi(\x(k)|\x(k))= \Phi (\x(k))$.
    \end{enumerate}
    The initial point $\x(0)\in\mathcal X$.
    
\end{defn}


For a non concave function, this MM algorithm converges to a stationary point~\cite{bolte2016majorization}. Convergence to global maximum can be ensured under further assumptions.

\begin{asu}\label{asu:str-ccv-surrogate}
    The minorizer $\phi(\x|\y)$ is differentiable and strongly concave in $\x$, with modulus $\beta>0$, and its gradient $\nabla\phi(\cdot|\y)$ is $B$-Lipschitz, uniformly over $\y$. Moreover, for every $\y\in\mathcal X$, the minorizer is tangent to the potential at $\y$, i.e., $\nabla\phi(\y|\y)=\nabla\Phi(\y)$.
\end{asu}

\begin{thm}\label{thm:convergence-thm}
    Under Assumptions \ref{asu:PL-condition} and \ref{asu:str-ccv-surrogate}, the MM learning scheme converges to a potential maximizing Nash equilibrium. Moreover, with $\tilde L\triangleq\max(B,\bar L)$, the optimality gap
    \begin{align}
        g(\x(k))\le \left(1-\frac{\mu}{\tilde L}\right)^{k}g(\x(0)).
    \end{align}
\end{thm}
\begin{proof}
    Since the minorizer is strongly concave with modulus $\beta$, we have, for every $\uv$, $\w$,
    \begin{align}\label{eqn:str-ccv-prop}
        \phi(\uv|\y)-\phi(\w|\y)\le \langle\uv-\w,\nabla \phi(\w|\y)\rangle-\frac{\beta}{2}||\uv-\w||^2.
    \end{align}
    Since $\x(k+1)$ maximizes the concave function $\phi(\cdot|\x(k))$ over the convex set $\mathcal X$, it satisfies the first order optimality condition $\langle\uv-\x(k+1),\nabla \phi(\x(k+1)|\x(k))\rangle\le 0$ for all $\uv\in\mathcal X$. Using \eqref{eqn:str-ccv-prop} with $\uv=\x(k)$, $\w=\x(k+1)$ and $\y=\x(k)$, we obtain
    \begin{align*}
        \phi(\x(k)|\x(k))+\frac{\beta}{2}||\x(k)-\x(k+1)||^2\le \phi(\x(k+1)|\x(k)).
    \end{align*}
    This implies,
    \begin{align*}
        \Phi (\x(k+1)) &\ge \phi(\x(k+1)|\x(k)),\\
        &\ge \phi(\x(k)|\x(k))+\frac{\beta}{2}||\x(k)-\x(k+1)||^2,\\
        &=\Phi (\x(k))+\frac{\beta}{2}||\x(k)-\x(k+1)||^2.
    \end{align*}
Hence,
\begin{align}
    g(\x(k))-g(\x(k+1))
    &\ge \frac{\beta}{2}||\x(k+1)-\x(k)||^2.\label{eqn:g-g-beta}
\end{align} 
 Following the argument of~\cite[Theorem~5]{karimi2016linearArxiv}, fix $\y\in\mathcal X$. Then,
\begin{align}
    \phi(\x|\y) &\ge \phi(\y|\y)+\langle\nabla\phi(\y|\y),\x-\y\rangle-\frac{B}{2}||\x-\y||^2,\\
    &\ge \Phi(\y)+\langle\nabla\Phi(\y),\x-\y\rangle-\frac{\tilde L}{2}||\x-\y||^2 \label{eqn:tangent-lower-bound}
\end{align}
where we used $\phi(\y|\y)=\Phi(\y)$ and Assumption \ref{asu:str-ccv-surrogate}. Taking $\y=\x(k)$ and maximizing both sides over $\x\in\mathcal X$, we obtain, using the definition \eqref{eqn:prox-PL-D},
\begin{align}
    \Phi (\x(k+1)) &\ge \phi(\x(k+1)|\x(k))=\max_{\x\in\mathcal X}\phi(\x|\x(k)),\\
    &\ge \Phi(\x(k))+\frac{1}{2\tilde L}\mathcal D_{\tilde L}(\x(k)).\label{eqn:max-minorizer-bound}
\end{align}
Using Lemma \ref{lem:D-increasing-property} and Assumption \ref{asu:PL-condition}, we get
\begin{align}
    g(\x(k))-g(\x(k+1)) &=\Phi (\x(k+1))-\Phi (\x(k)),\\
    &\ge \frac{\mu}{\tilde L}\, g(\x(k)).\label{eqn:g-iterate-rel}
\end{align}
This yields $g(\x(k+1))\le (1-\frac{\mu}{\tilde L})g(\x(k))$, where $\mu\le \tilde L$ since $g\ge 0$, and therefore
\begin{align}
    g(\x(k))\le \left(1-\frac{\mu}{\tilde L}\right)^{k}g(\x(0)).
\end{align}
As $k\to\infty$, $g(\x(k))\to 0$, and hence,
$
    \Phi (\x(k))\to\Phi^*.
$
From \eqref{eqn:g-g-beta} and using the fact that $g(\x(k))$ is non negative and decreasing, we can write
\begin{align}
   ||\x(k+1)-\x(k)|| &\le \sqrt{\frac{2}{\beta}g(\x(k))},\\
   &\le \sqrt{\frac{2g(\x(0))}{\beta}} \left(1-\frac{\mu}{\tilde L}\right)^{\frac{k}{2}}.
\end{align}
Thus $(\x(k))_{k\ge 0}$ is Cauchy, and converges to some $\hat\x$. Since $\Phi (\x(k))$ converges to $\Phi^*$, it follows that $\hat\x$ is a maximizer of the potential function $\Phi (\x)$.
\end{proof}
\section{Privacy Aware MM}\label{sec:Privacy-preserving-MM}
While the minorizer $\phi$ has to be designed keeping the requirements of Theorem \ref{thm:convergence-thm} in mind, maximization of the minorizer (which is to be done in every iteration of the MM algorithm), can itself be implemented in a partially distributed fashion. The need for a distributed implementation stems from the fact that agents' individual reward functions encode information about their personal preferences through their prospect-theoretic value functions, which they may be reluctant to disclose to other agents.

A minorizer that enables distributed optimization at each iteration, while limiting the disclosure of agents' preference information to one another, is obtained by keeping the collective term $\J$ exact and linearizing only the individual terms. Write the potential as
\begin{align}
    \Phi(\x)=\J(\x)+\sum_{i=1}^N f_i(x_i),
\end{align}
with $f_i(x_i)\triangleq\frac{1}{a_i}\mathbb E_{\sigma}[V_i\circ\mathcal R_i(x_i,\xi)]$,
and consider
\begin{align}\label{eqn:partial-minorizer}
    \phi(\x|\y)=\J(\x)+\sum_{i=1}^N h_i(x_i,y_i),
\end{align}
where
\begin{align}\label{eqn:h-i-def}
    h_i(x_i,y_i)=f_i(y_i)+f_i'(y_i)(x_i-y_i)-\frac{\alpha}{2}(x_i-y_i)^2.
\end{align}
If each $f_i'$ is $\alpha$-Lipschitz, the descent lemma~\cite[Lemma 1.2.3]{nesterov2004introductory} gives $h_i(x_i,y_i)\le f_i(x_i)$, with equality at $x_i=y_i$, so that \eqref{eqn:partial-minorizer} is a minorizer of $\Phi$. Since $\nabla\phi(\y|\y)=\nabla\J(\y)+(f_1'(y_1),\dots,f_N'(y_N))=\nabla\Phi(\y)$, it is tangent to $\Phi$, and if $\J$ is concave with $L_{\J}$-Lipschitz gradient, it satisfies Assumption \ref{asu:str-ccv-surrogate} with $\beta\ge\alpha$ and $B=\alpha+L_{\J}$. Thus, there is a separable structure in the minorizer: the collective term $\J$ couples the agents, while each $h_i$ depends only on the strategy and on the PT preferences of agent $i$. If $\J$ is a separable function, then the optimization becomes completely separable, and can be solved in a distributed manner. Linearizing the whole potential instead would make the MM step a projected gradient step~\cite{karimi2016linearArxiv}, but would not preserve this separable structure.
\subsection{Noisy Alternating Direction Method of Multipliers (ADMM) Method}
An ADMM based approach can be used to maximize \eqref{eqn:partial-minorizer} if the functions $\J$ and $h_i$ are concave. To limit the information revealed to other agents, each agent can inject noise into its individual updates, thereby making its state variables less directly observable.

For a concave function $f$, define the proximal operator,
\begin{align}
    \mathbf{prox}_{\e f}(\uv)=\arg_{\x}\max\left(f(\x)-\frac{1}{2\e}||\x-\uv||^2\right).
\end{align}
The ADMM algorithm~\cite{parikh2014proximal} is
\begin{align}
    \x^{t+1} &=\mathbf{prox}_{\e \J}(\z^t-\vu^t),\label{eqn:admm-x}\\
    z_i^{t+1} &=\mathbf{prox}_{\e h_i}(x_i^{t+1}+v_i^t)+\eta_i^{t+1},~i=1,...,N,\\
        \vu^{t+1} &= \vu^t+\x^{t+1}-\z^{t+1}.\label{eqn:admm-v}
\end{align}
The first step is generally centralized (unless $\J$ is separable) and is performed by agent $N+1$. The other two steps can be carried out by the agents in a distributed fashion. Importantly, the individual prospect-theoretic value functions $f_i$ encode agents' personal preferences, through parameters that characterize how they evaluate outcomes under risk. Their derivatives $f_i'(y_i)$, which enter the individual updates, may therefore reveal information about these preferences. To limit such disclosure, agents inject noise $\eta_i$ into their updates, so that the messages sent to the coordinator do not reveal the exact values of $f_i'(y_i)$. As shown in Section~\ref{sec:inexact-MM}, convergence to $\Phi^*$ requires the variance of this noise to decrease over the MM iterations.

\section{Convergence with an Inexact MM Step}\label{sec:inexact-MM}

The noisy ADMM of
Section~\ref{sec:Privacy-preserving-MM} can be considered an MM scheme where the optimization in the MM step returns an answer with error. We first show that MM can tolerate certain errors and then show that the noisy ADMM
produces errors of the required kind, under assumptions on the collective benefit $\J$.

Throughout, $\tilde L=\max(B,\bar L)$, and
\begin{align}
    \zeta(\y)\triangleq\arg_{\x\in\mathcal X}\max\phi(\x|\y)
\end{align}
denotes the exact maximization step in the MM algorithm. From Assumption \ref{asu:str-ccv-surrogate}, it follows that this is unique.

\subsection{MM maximization with error}

The error of the inexact MM step is 
\begin{align}
    \mathbf e_k\triangleq \x(k+1)-\zeta(\x(k)).
\end{align}
Let $\mathcal F_k$ denote the history of the algorithm up to $\x(k)$. We also
define
\begin{align}\label{eqn:G-def}
    G\triangleq\sup_{\y\in\mathcal X}||\nabla\phi(\zeta(\y)|\y)||.
\end{align}
By the $B$-Lipschitz gradient and the tangency of $\phi$ (Assumption~\ref{asu:str-ccv-surrogate}), $||\nabla\phi(\zeta(\y)|\y)||\le||\nabla\Phi(\y)||+B||\zeta(\y)-\y||$, so that $G\le\sup_{\x\in\mathcal X}||\nabla\Phi(\x)||+B\,\mathrm{diam}(\mathcal X)<\infty$.
Moreover, $G=0$ when $\zeta(\y)$ lies in the interior of $\mathcal X$ for
every $\y$, since $\nabla\phi(\zeta(\y)|\y)=0$ at an interior maximizer.

\begin{thm}[MM with error]\label{thm:inexact}
    Let Assumptions \ref{asu:PL-condition} and \ref{asu:str-ccv-surrogate} hold, let $\rho\triangleq 1-\frac{\mu}{\tilde L}$, and suppose that $\left(\mathbb E\left[||\mathbf e_k||^2\,\big|\,\mathcal F_k\right]\right)^{1/2}\le\varepsilon_k$ for a deterministic sequence $(\varepsilon_k)$. Then
    \begin{align}\label{eqn:inexact-exp}
        \mathbb E\, g(\x(k))\le\rho^{k}g(\x(0))+\sum_{j=0}^{k-1}\rho^{k-1-j}\left(G\varepsilon_j+\frac{B}{2}\varepsilon_j^2\right),
    \end{align}
    and, with $\bar\varepsilon\triangleq\limsup_k\varepsilon_k$,
    \begin{align}\label{eqn:inexact-limsup}
        \limsup_{k\to\infty}\mathbb E\, g(\x(k))\le\frac{\tilde L}{\mu}\left(G\bar\varepsilon+\frac{B}{2}\bar\varepsilon^2\right).
    \end{align}
    In particular, $\mathbb E\, g(\x(k))\to 0$ if $\varepsilon_k\to 0$.
\end{thm}
\begin{proof}
    Fix $k$, and write $\y=\x(k)$ and $\hat\x=\zeta(\y)$, so that $\x(k+1)=\hat\x+\mathbf e_k$. By the minorization property and the $B$-Lipschitz gradient of $\phi(\cdot|\y)$ (Assumption~\ref{asu:str-ccv-surrogate}),
    \begin{align*}
        \Phi(\x(k+1))&\ge\phi(\x(k+1)|\y)\\
        &\ge\phi(\hat\x|\y)+\langle\nabla\phi(\hat\x|\y),\mathbf e_k\rangle-\frac{B}{2}||\mathbf e_k||^2\\
        &\ge\phi(\hat\x|\y)-G||\mathbf e_k||-\frac{B}{2}||\mathbf e_k||^2 .
    \end{align*}
    As in \eqref{eqn:max-minorizer-bound}, by Lemma~\ref{lem:D-increasing-property} and Assumption~\ref{asu:PL-condition}, $\phi(\hat\x|\y)\ge\Phi(\y)+\frac{\mu}{\tilde L}g(\y)$, and subtracting both sides from $\Phi^*$ gives
    \begin{align}\label{eqn:inexact-rec}
        g(\x(k+1))\le\rho\, g(\x(k))+G||\mathbf e_k||+\frac{B}{2}||\mathbf e_k||^2 .
    \end{align}
    Taking conditional expectations, with $\mathbb E[||\mathbf e_k||\,|\,\mathcal F_k]\le\varepsilon_k$ by Jensen's inequality, and unrolling the recursion gives \eqref{eqn:inexact-exp}. Finally, the weights $\rho^{k-1-j}$ in \eqref{eqn:inexact-exp} sum to at most $1/(1-\rho)=\tilde L/\mu$, which gives \eqref{eqn:inexact-limsup}.
\end{proof}

    Theorem~\ref{thm:inexact} makes no assumption on how the MM step is solved: the errors need not be zero-mean or independent, and only their conditional second moments are used. The errors enter \eqref{eqn:inexact-exp} through two terms. The linear term is present only when $\zeta(\y)$ lies on the boundary of $\mathcal X$. When $\zeta(\y)$ is interior for all $\y$, $G=0$ and the errors have a quadratic effect, so that the asymptotic gap scales with the variance of the error. Finally, since $\Phi$ is continuous on the compact set $\mathcal X$, $g(\x(k))\to 0$ implies that the distance from $\x(k)$ to the set of maximizers $\mathbf M$ tends to zero; when $\Phi$ has a unique maximizer $\x^*$, $\x(k)\to\x^*$.

\subsection{Noisy ADMM with finite iterations}

We now consider the case where the MM step is solved by the noisy ADMM of
Section~\ref{sec:Privacy-preserving-MM}. At MM step $k$, with $\y=\x(k)$, the
algorithm runs $T$ iterations of \eqref{eqn:admm-x}--\eqref{eqn:admm-v}, with $h_i=h_i(\cdot,y_i)$, from $\z^0=\y$ and $\vu^0=0$, and sets
$\x(k+1)=\Pi_{\mathcal X}(\x^T)$, the projection of the last iterate onto
$\mathcal X$. With the state $\w=(\z,\vu)$, the noiseless iteration
($\eta\equiv 0$) is a map $\w^{t+1}=F_{\y}(\w^t)$, and the noisy one reads
\begin{align}\label{eqn:noisy-admm-state}
    \w^{t+1}=F_{\y}(\w^t)+(\eta^{t+1},-\eta^{t+1}),
\end{align}
since the noise added to $\z^{t+1}$ is subtracted in the update of $\vu^{t+1}$.

\begin{lem}[Error of noisy ADMM]\label{lem:admm}
    Let the minorizer be \eqref{eqn:partial-minorizer}, with $\alpha>0$ and $\J(\x)=-\frac{1}{2}\x^\top Q\x+\mathbf p^\top\x$ for some $Q\succeq 0$ and $\mathbf p\in\mathbb R^N$, and suppose that, for every $\y\in\mathcal X$, the maximizer of $\phi(\cdot|\y)$ over $\mathbb R^N$ lies in $\mathcal X$. If $\mathbb E\left[||\eta^{t}||^2\,\big|\,\mathcal F_k\right]\le N\nu_k$ for $t=1,\dots,T$, then there are constants $\bar\rho\in(0,1)$ and $C$, independent of $k$, such that
    \begin{align}\label{eqn:admm-err}
        \left(\mathbb E\left[||\mathbf e_k||^2\,\big|\,\mathcal F_k\right]\right)^{1/2}\le\varepsilon_k\triangleq C\left(\bar\rho^{\,T}+\sqrt{N\nu_k}\right).
    \end{align}
\end{lem}
\begin{proof}
    Since $\J$ and $h_i(\cdot,y_i)$ are quadratic, the noiseless iteration is affine, $F_{\y}(\w)=M\w+b_{\y}$: the linear parts of the proximal operators, $(I+\e Q)^{-1}$ and $(1+\e\alpha)^{-1}$, do not depend on $\y$, while $b_{\y}$ depends continuously on $\y$ through $f_i'(y_i)$ and $y_i$. As the noiseless ADMM converges linearly to the maximizer $\zeta(\y)$ of $\phi(\cdot|\y)$~\cite{giselsson2016linear,deng2016global}, $||M^t||\le C_M\bar\rho^{\,t}$ for some $\bar\rho<1$, and $F_{\y}$ has a unique fixed point $\w^\star_{\y}=(\z^\star_{\y},\vu^\star_{\y})$, with $\z^\star_{\y}=\mathbf{prox}_{\e\J}(\z^\star_{\y}-\vu^\star_{\y})=\zeta(\y)$. Unrolling \eqref{eqn:noisy-admm-state},
    \begin{align*}
        \w^{T-1}-\w^\star_{\y}&=M^{T-1}(\w^{0}-\w^\star_{\y})\\
        &\quad+\sum_{s=1}^{T-1}M^{T-1-s}(\eta^{s},-\eta^{s}).
    \end{align*}
    Since the proximal operator and the projection onto $\mathcal X$ are nonexpansive, $||\mathbf e_k||\le\sqrt{2}\,||\w^{T-1}-\w^\star_{\y}||$. By Minkowski's inequality, using $\mathbb E\left[||(\eta^{s},-\eta^{s})||^2\,\big|\,\mathcal F_k\right]\le 2N\nu_k$,
    \begin{align*}
        &\left(\mathbb E\left[||\mathbf e_k||^2\,\big|\,\mathcal F_k\right]\right)^{1/2}\\
        &\quad\le\sqrt{2}\,C_M\left(\bar\rho^{\,T-1}||\w^{0}-\w^\star_{\y}||+\frac{\sqrt{2N\nu_k}}{1-\bar\rho}\right),
    \end{align*}
    which gives \eqref{eqn:admm-err}, since $||\w^{0}-\w^\star_{\y}||$ is bounded over the compact set $\mathcal X$.
\end{proof}

    After $T$ inner iterations, the initial error has shrunk as $\bar\rho^{\,T}$, while the accumulated noise remains of order $\sqrt{N\nu_k}$. The interior assumption is needed because the ADMM of Section~\ref{sec:Privacy-preserving-MM} does not take the constraints into account. Otherwise, each agent can project its update onto $\mathcal X_i$, and a similar bound follows from the contraction property of Douglas--Rachford splitting~\cite{giselsson2016linear}.

\begin{cor}[MM with noisy ADMM]\label{cor:noisy-admm}
    Under the assumptions of Theorem~\ref{thm:inexact} and Lemma~\ref{lem:admm}, the bound \eqref{eqn:inexact-exp} holds with $\varepsilon_k$ given by \eqref{eqn:admm-err}. In particular, if $\nu_k\to 0$, 
    \begin{align}\label{eqn:noisy-admm-floor}
        \limsup_{k\to\infty}\mathbb E\left[g(\x(k))\right]\le\frac{\tilde L}{\mu}\left(G\bar\varepsilon+\frac{B}{2}\bar\varepsilon^2\right),
    \end{align}
    where $\bar\varepsilon\triangleq C\bar\rho^{\,T}$.
\end{cor}
\begin{proof}
    By Lemma~\ref{lem:admm}, Theorem~\ref{thm:inexact} applies with $\varepsilon_k$ given by \eqref{eqn:admm-err}. If $\nu_k\to 0$, then $\varepsilon_k\to\bar\varepsilon$, and \eqref{eqn:inexact-limsup} gives \eqref{eqn:noisy-admm-floor}.
\end{proof}

    The bound of Corollary~\ref{cor:noisy-admm} separates three effects: the initial gap is forgotten at the rate $\rho$ of Theorem~\ref{thm:convergence-thm}; the noise contributes through $\sqrt{\nu_k}$ and vanishes only if its variance decreases to zero, independent and identically distributed noise leaving an asymptotic gap; and the finite number $T$ of inner iterations leaves the floor \eqref{eqn:noisy-admm-floor}, which decreases exponentially in $T$ and vanishes if $T$ grows with $k$.

    The injected noise only obscures the agents' individual updates. Since its variance must decrease for the iterates to converge, the level of obfuscation it provides also diminishes over time. This highlights a trade-off between convergence accuracy and privacy protection in noise-based mechanisms: for
    schemes that perturb the exchanged messages, exact convergence and differential privacy cannot be achieved  simultaneously~\cite{nozari2018differentially}. Larger noise levels may provide stronger obfuscation of individual preference information, but can also increase the error in the MM updates and slow down convergence. Conversely, reducing the noise improves solution accuracy and convergence, but progressively weakens the protection of individual information.

\section{Numerical Example}\label{sec:numericals}
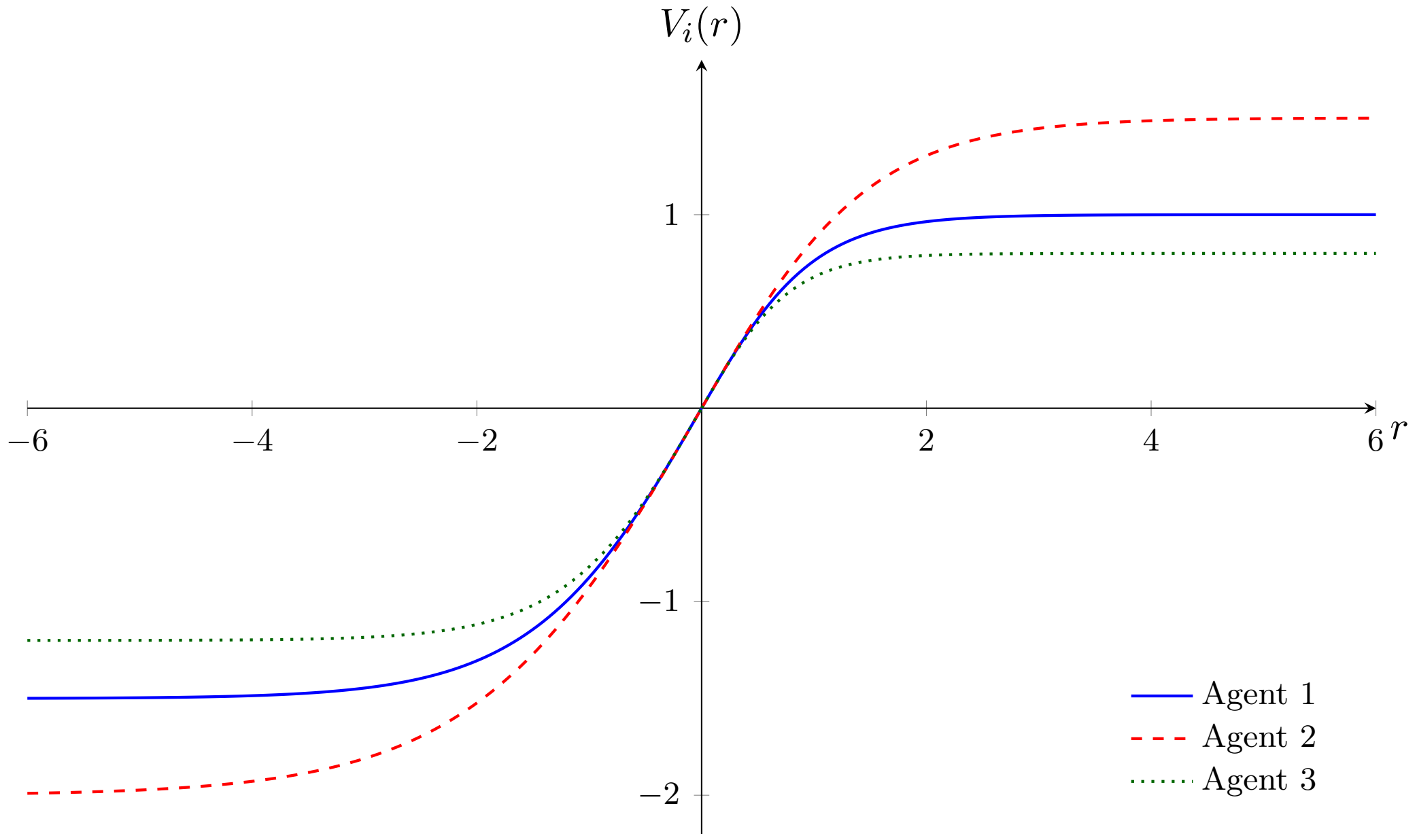
\begin{figure}[t]
\centering
\begin{tikzpicture}[scale=1]
\begin{axis}[
  width=\columnwidth, height=0.62\columnwidth,
  xmin=-6, xmax=6, ymin=-2.2, ymax=1.8,
  axis lines=middle, samples=120,
  xlabel={$r$}, ylabel={$V_i(r)$},
  x label style={anchor=north west}, y label style={anchor=south},
  xtick={-6,-4,-2,2,4,6}, ytick={-2,-1,1},
  tick label style={font=\footnotesize},
  legend pos=south east, legend cell align=left,
  legend style={font=\footnotesize, draw=none, fill=none},
]
\addplot[thick, blue, domain=0:6] {1.0*tanh(x/1.0)};
\addlegendentry{Agent 1}
\addplot[thick, blue, domain=-6:0, forget plot] {1.5*tanh(x/1.5)};
\addplot[thick, red, dashed, domain=0:6] {1.5*tanh(x/1.5)};
\addlegendentry{Agent 2}
\addplot[thick, red, dashed, domain=-6:0, forget plot] {2.0*tanh(x/2.0)};
\addplot[thick, black!60!green, dotted, domain=0:6] {0.8*tanh(x/0.8)};
\addlegendentry{Agent 3}
\addplot[thick, black!60!green, dotted, domain=-6:0, forget plot] {1.2*tanh(x/1.2)};
\end{axis}
\end{tikzpicture}
\caption{Prospect-theoretic value functions $V_i$ used in the example:
concave on gains ($r>0$) with saturation level $\theta_i^{+}$, convex on
losses ($r<0$) with saturation level $-\theta_i^{-}$, where
$\theta_i^{-} > \theta_i^{+}$.}
\label{fig:value}
\end{figure}

We illustrate our results on an example which captures the key features
of our analysis: agents with smooth S-shaped value functions that make the potential non-concave while satisfying
our assumptions; a collective benefit that couples the players; a game small enough for the optimum to be computed exactly. We consider $N=3$ agents and a coordinator (agent $N+1$). Agent $i$ chooses
$x_i \in \mathcal{X}_i = [0,\bar{x}]$, interpreted as a quantity sold at a random
price $\xi \in \{0.5,\,1,\,2\}$ with probabilities $\q=(0.3,\,0.5,\,0.2)$. The
reward of agent $i$ is $\mathcal R_i(x_i,\xi) = \xi x_i - b_i$, where $b_i>0$ is a
reference level. Outcomes are evaluated through the smooth
prospect-theoretic value function
\begin{equation}
V_i(r) =
\begin{cases}
\theta_i^{+}\tanh\!\left(r/\theta_i^{+}\right), & r \ge 0,\\[2pt]
\theta_i^{-}\tanh\!\left(r/\theta_i^{-}\right), & r < 0,
\end{cases}
\end{equation}
which is increasing, concave on gains and convex on losses 
(Fig.~\ref{fig:value}).

 The
weights $\sigma$ are obtained from $\q$ via \eqref{eqn:q-tilde-gen} with
$\pi(p) = p^{\delta}/\big(p^{\delta}+(1-p)^{\delta}\big)^{1/\delta}$,
$\delta = 0.65$, giving $\sigma = (0.324,\,0.316,\,0.360)$. The collective
benefit is
\begin{align}
\mathcal{J}(\x) &= -\frac{\kappa}{2}\sum_{i=1}^{N}(x_i-d)^2
-\frac{\gamma}{2}\sum_{i<j}(x_i-x_j)^2.
\end{align}
Its first
part penalizes, quadratically, the deviation of each agent from its share $d$
of a collective target (as in tracking or imbalance penalties), while the second
part penalizes disparities
between agents and thus favours an equitable distribution of effort. It can also be written as
\begin{align}
 \J(\x) = -\frac{1}{2}(\x-d\mathbf{1})^{\top} Q\,(\x-d\mathbf{1}),   
\end{align}
with $Q=\kappa I+\gamma \Lambda$, where $\Lambda$ is the Laplacian of the complete graph. This form
is also chosen for tractability of the coordinator's proximal step in the ADMM. The utilities are
$J_i(\x) = a_i\mathcal{J}(\x) + \mathbb{E}_{\sigma}[V_i\circ \mathcal R_i(x_i,\xi)]$.
The different parameters are listed in Table~\ref{tab:params}.

\begin{table}[h]
\centering
\caption{Parameters of the numerical example.}
\label{tab:params}
\begin{tabular}{lccc}
\toprule
 & Agent 1 & Agent 2 & Agent 3 \\
\midrule
$a_i$          & 1.0 & 0.8 & 0.5 \\
$b_i$          & 2.0 & 1.5 & 2.5 \\
$\theta_i^{+}$ & 1.0 & 1.5 & 0.8 \\
$\theta_i^{-}$ & 1.5 & 2.0 & 1.2 \\
\midrule
\multicolumn{4}{l}{$\kappa = 0.2,\quad \gamma = 0.1,\quad d = 5,\quad \bar{x} = 17.58$} \\
\bottomrule
\end{tabular}
\end{table}

When $\xi_1 d \ge \max_i b_i$, one
can  show that the potential $\Phi$ has a unique critical point, which is
its global maximizer $\x^* = (5.617,\,5.651,\,5.913)$, lying in the interior of
$\mathcal{X}$, and that $\Phi$ satisfies Assumption~\ref{asu:PL-condition} (numerically,
$\mu \approx 0.18$ for any $\bar L\ge 1$). Nevertheless, $\Phi$ and each $J_i$ are non-concave:
$\partial^2 \Phi/\partial x_i^2 > 0$  near $x_i = 0$.

\paragraph{Algorithm settings}
At each MM step the minorizer \eqref{eqn:partial-minorizer} of
Section~\ref{sec:Privacy-preserving-MM},
$\phi(\x|\y) = \mathcal{J}(\x) + \sum_{i} h_i(x_i,y_i)$, with
\begin{equation}
h_i(x_i,y_i) = f_i(y_i) + f_i'(y_i)(x_i-y_i) - \frac{\alpha}{2}(x_i-y_i)^2
\end{equation}
and $f_i(x_i) = a_i^{-1}\mathbb{E}_{\sigma}[V_i\circ \mathcal R_i(x_i,\xi)]$, is
maximized by $30$ iterations of noisy ADMM with $\lambda = 1$. Note that $\phi$ is a minorizer of
$\Phi$ whenever $\alpha \ge \sup_x |f_i''(x)|$ for all $i$. We take $\alpha$
equal to the closed-form bound obtained from $|V_i''| \le 4/(3\sqrt{3}\,\theta)$
with $\theta = \min(\theta_i^{+},\theta_i^{-})$,
\begin{equation}
\alpha = \max_i \frac{1}{a_i}\sum_{m}\sigma_m\xi_m^2\,
\frac{4}{3\sqrt{3}\,\min(\theta_i^{+},\theta_i^{-})} = 3.535,
\end{equation}
which guarantees minorization and can be computed without any numerical search. At MM step $k$, agents add independent noise
$\eta_i \sim \mathrm{Laplace}(0, s_k)$ with $s_k = s_0\,(0.95)^k$, for initial
variances $\mathrm{Var}(\eta_0) = 2s_0^2 \in \{0,\,0.01,\,0.1,\,1,\,5\}$. Each
configuration is run $20$ times from $\x(0) = \mathbf{0}$ for $150$ MM steps, and the
reported curves are averages over runs.

\begin{figure}
\centering
\includegraphics[scale=1]{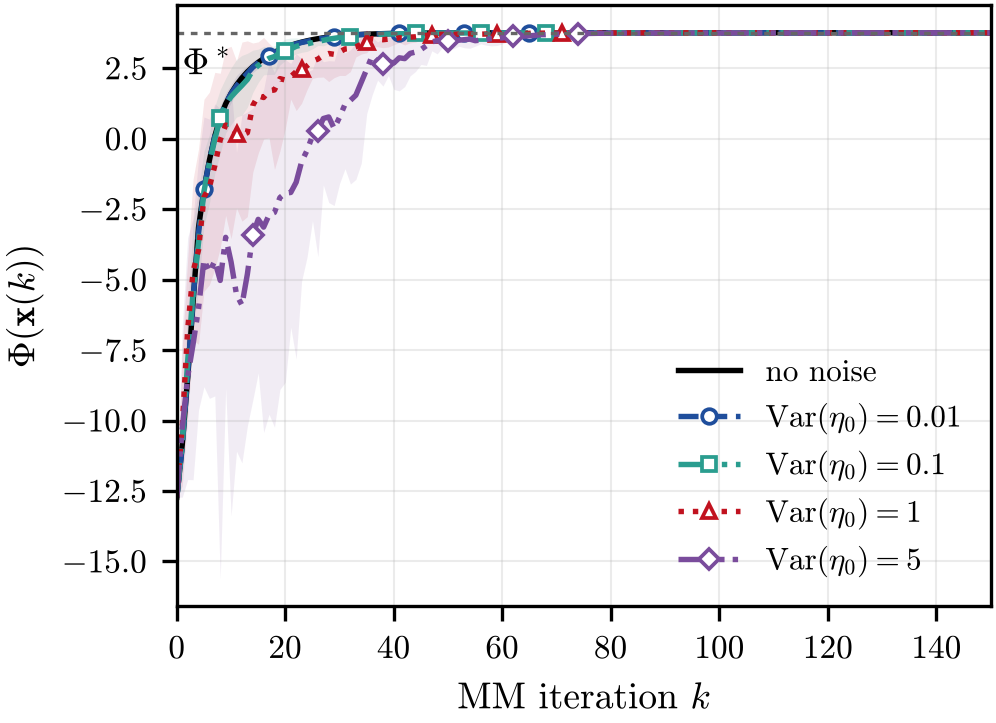}
\caption{Potential $\Phi(\x(k))$ along the MM iterates for different initial
noise variances, averaged over $20$ runs; shaded regions show the
$10$--$90\%$ range across runs. The dashed line is $\Phi^*$.}
\label{fig:potential}
\end{figure}

\begin{figure}
\centering
\includegraphics[scale=1]{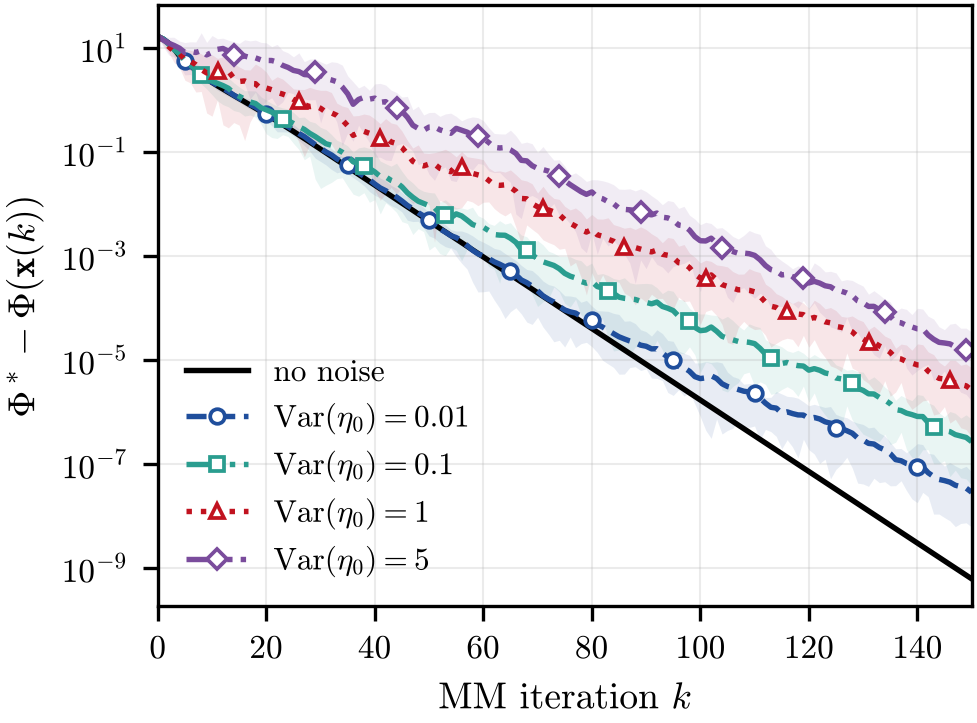}
\caption{Optimality gap $\Phi^* - \Phi(\x(k))$ (log scale) for different
initial noise variances, averaged over $20$ runs; shaded regions show the
$10$--$90\%$ range across runs.}
\label{fig:gap}
\end{figure}

\paragraph{Results}
Fig.~\ref{fig:potential} shows the potential $\Phi(\x(k))$ along the MM
iterates, and Fig.~\ref{fig:gap} the optimality gap $\Phi^* - \Phi(\x(k))$ on
a logarithmic scale. For every noise level the iterates converge to the
potential maximizing equilibrium $\x^*$. A larger initial variance slows the
early iterations, where the noise is largest, but since $s_k \to 0$ it does
not leave a residual gap: after $150$ MM steps the mean gap ranges from
$6\times 10^{-10}$ without noise to $1.3\times 10^{-5}$ for
$\mathrm{Var}(\eta_0) = 5$. With noise, the gap decays geometrically at a rate
governed by the noise decay $0.95^k$ rather than by the MM contraction alone. This is the behaviour predicted by Corollary~\ref{cor:noisy-admm}: in
this example the exact MM step is interior, so that $G=0$, and since the noise
variance decays to zero, the only remaining floor is due to the finite number of
ADMM iterations, which is negligible for $T=30$.

\begin{figure}
\centering
\includegraphics[scale=1]{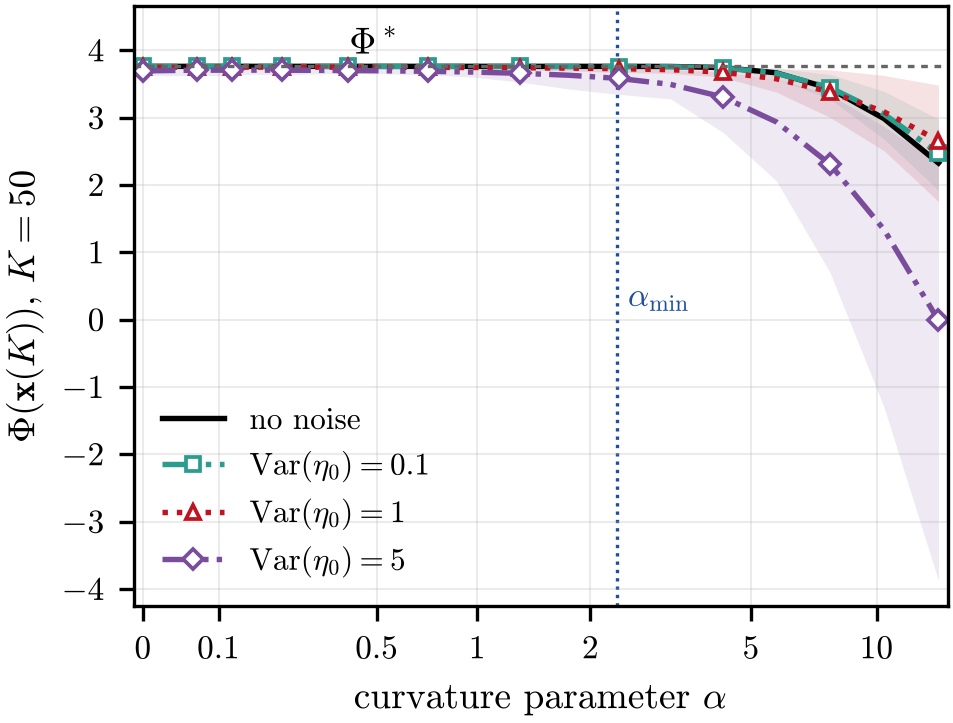}
\caption{Potential $\Phi(\x(K))$ after $K = 50$ MM steps versus the curvature
parameter $\alpha$, for different initial noise variances, averaged over $20$
runs; shaded regions show the $10$--$90\%$ range across runs. The dotted
line marks $\alpha = \alpha_{\min}$, and the dashed line is $\Phi^*$. The horizontal axis is
logarithmic in $\alpha$ plus a small constant (0.2), so that $\alpha = 0$ can be shown.}
\label{fig:final_alpha}
\end{figure}

\paragraph{Impact of prospect theory}
If we consider the same game with expected-utility
players, i.e., $V_i(r)=r$ and $\sigma=\q$, the potential is a concave quadratic, and the equilibrium is
$\x_{\mathrm{EU}}=(11.56,\,12.09,\,13.66)$, roughly twice the PT equilibrium
$\x^*=(5.62,\,5.65,\,5.91)$. The difference is driven by the value function: at equilibrium all outcomes are gains, where $V_i$ is concave and saturates, so the PT players, being risk averse in gains, sell much
smaller quantities. Probability weighting alone has the opposite effect: using $\sigma$ with
$V_i(r)=r$ gives $(12.49,\,13.09,\,14.88)$. This is because the weighting overweights the rare high price ($\sigma_3=0.36$
against $q_3=0.2$), which raises the perceived expected price
($\mathbb E_\sigma[\xi]=1.20$ against $\mathbb E_{\q}[\xi]=1.05$). Also note that  PT is also what makes the problem
non-concave: the non-concavity of $\Phi$ comes from the convex loss region of
$V_i$, and is absent under expected utility, where standard methods apply.
\paragraph{Effect of the minorizer curvature}
The bound $\alpha = 3.535$ used above is conservative. For the minorizer
\eqref{eqn:partial-minorizer}, the condition of Section~\ref{sec:Privacy-preserving-MM}
($f_i'$ $\alpha$-Lipschitz) holds for $\alpha \ge \max_i\sup_{x\in\mathcal X_i}|f_i''(x)| = 2.34$,
which here coincides with
$\alpha_{\min} \triangleq \max_i \sup_{x\in\mathcal{X}_i}\big(-f_i''(x)\big)$, the smallest $\alpha$
for which $\phi$ is a minorizer; for $\alpha<\alpha_{\min}$ the minorization is
imperfect. To study the
effect of $\alpha$, we run the full algorithm for a fixed budget of $K = 50$ MM
steps, for $21$ values of $\alpha \in [0, 14]$ and the same decaying noise as
above ($20$ runs per configuration, $150$ ADMM iterations per MM step).
Fig.~\ref{fig:final_alpha} shows the resulting value $\Phi(\x(K))$, with the
dotted line marking $\alpha = \alpha_{\min}$.

To the right of $\alpha_{\min}$, a larger $\alpha$ makes $\phi$ more curved and each MM
step shorter, so after $K$ steps the iterates are further from the optimum: at
$\alpha = 14$ the mean of $\Phi(\x(K))$ is about $2.3$ without noise and about
$0$ for $\mathrm{Var}(\eta_0) = 5$, against $\Phi^* = 3.76$, and the noise also
widens the spread across runs. To the left of $\alpha_{\min}$, where $\phi$ is no
longer a minorizer and Theorem~\ref{thm:convergence-thm} does not apply, the curves
stay close to $\Phi^*$: in all runs the algorithm still converged to $\x^*$. In this
region the remaining gap is set by the noise rather than by $\alpha$; it is below
$10^{-5}$ without noise and at most $3\times 10^{-3}$, $3\times 10^{-2}$ and $0.13$ for
$\mathrm{Var}(\eta_0) = 0.1$, $1$ and $5$. Choosing $\alpha$ above $\alpha_{\min}$ thus
 slows convergence, while values below $\alpha_{\min}$ can be faster but are
not covered by Theorem \ref{thm:convergence-thm}.

\section{Conclusion}
We studied learning in non-cooperative potential games with prospect theoretic non concave utilities. If the potential satisfies a proximal P\L{} inequality, an MM learning algorithm can be constructed, which converges linearly to a potential maximizing Nash equilibrium. The minorizer for this algorithm can be designed to enable distributed implementation using a noisy ADMM. The noise prevents the agents from revealing their preferences exactly. Convergence of the noisy MM algorithm is shown under assumptions of the noise variance. In particular, we demonstrate a trade off between algorithm convergence and privacy for agent preferences:  noise levels that do not decay in time ensure better hiding of agent preferences, whereas decaying noise levels lead to convergence to equilibrium.  Numerical results illustrate the performance of the proposed algorithm. A formal characterization of the privacy guarantee will be undertaken in a follow-up work.

\subsection*{Declaration of Generative AI and AI-Assisted Technologies}
The authors used Claude (Opus 5.5, Anthropic) to assist in checking and deriving intermediate algebraic steps and formatting mathematical derivations in Sections \ref{sec:PL-conv} and \ref{sec:inexact-MM}, for designing the numerical simulations in Section \ref{sec:numericals}, and for formatting and editing the paper. The authors have independently re-derived and verified all the content of the paper, and take full responsibility for the mathematical accuracy and content of the publication.

\bibliographystyle{ieeetr}
\bibliography{sample}
\end{document}